\documentclass[twocolumn,footinbib,prb,showpacs,preprintnumbers,amsmath,amssymb]{revtex4}

\usepackage{graphicx}
\usepackage{dcolumn}
\usepackage{bm}

\begin{document}

\preprint{}

\title{Dynamics of Quasi-ordered Structure in a Regio-regulated $\pi$-Conjugated Polymer:Poly(4-methylthiazole-2,5-diyl)}

\author{Seiichiro Mori}
\affiliation{Department of Biomolecular Engineering, Tokyo Institute of Technology.  4259 Nagatsuta-cho, Midori-ku, Yokohama, Kanagawa 226-8501, Japan}

\author{Takakazu Yamamoto}%
\affiliation{Chemical Resources Laboratory, Tokyo Institute of Technology.  4259 Nagatsuta-cho, Midori-ku, Yokohama, Kanagawa 226-8503, Japan}

\author{Yoshio Inoue}
\affiliation{Department of Biomolecular Engineering, Tokyo Institute of Technology. 4259 Nagatsuta-cho, Midori-ku, Yokohama, Kanagawa 226-8501, Japan}

\author{Naoki Asakawa}
\thanks{Corresponding Author}
\email{nasakawa@bio.titech.ac.jp}
\affiliation{Department of Biomolecular Engineering, Tokyo Institute of Technology.  4259 Nagatsuta-cho B-55, Midori-ku, Yokohama, Kanagawa 226-8501, Japan}%

\date{\today}

\begin{abstract}
Dynamics of regio-regulated
Poly(4-methylthiazole-2,5-diyl)[HH-P4MeTz] was
investigated by solid-state $^1$H, $^2$D, $^{13}$C NMR spectroscopies, and differential scanning calorimetry(DSC) measurements. 
DSC,
$^2$D quadrupolar echo NMR, $^{13}$C cross-polarization and magic-angle 
spinning(CPMAS) NMR, and 2D spin-echo(2DSE) CPMAS NMR spectroscopy
suggest existence of a quasi-ordered phase in which backbone twists
take place with weakened $\pi$-stackings.
Two-dimensional exchange $^2$D NMR(2DEX) detected slow dynamics with a rate of
an order of 10$^2$Hz for the CD$_3$ group in $\it d$$_3$-HH-P4MeTz at 288K.
The frequency dependence of proton longitudinal relaxation
rate at 288K shows
a $\omega^{-1/2}$ dependence,
which is due to the one-dimensional
diffusion-like motion of backbone conformational
modulation waves.
The diffusion rate was estimated as 3$\pm$2 GHz, which was approximately
10$^7$ times larger than that estimated by
2DEX NMR measurements.
These results suggest that there exists anomalous dispersion
of modulation waves in HH-P4MeTz.
The one-dimensional group velocity of the wave packet is responsible for the behavior of proton
longitudinal relaxation time.
On the other hand, the 2DEX NMR is sensitive to phase
velocity of the nutation of methyl groups that is associated with backbone twists.
From proton T$_1$ and T$_2$ measurements, the activation energy was estimated as 2.9 and 3.4 kcal/mol, respectively.
These were in agreement with 3.0 kcal/mol determined by M\o ller-Plesset(MP2) molecular orbital(MO) calculation.
We also performed chemical shielding calculation of the methyl-carbon in order to understand chemical shift tensor behavior, leading to the fact that a quasi-ordered phase coexist with the crystalline phase.
\end{abstract}

\maketitle

\section{Introduction}
In the discussion of structure-property correlation
in polymeric condensed materials such as semicrystalline
polymers and
polymeric liquid crystals,
one must often take into account of not only duality of
crystal/amorphous
but also elemental excitations of polymeric condensed
systems or
their relaxation from elementally excited states to a ground state.
For instance, exciton
generation and anihilation in $\pi$-conjugated
polymers
and dynamic disorder in polymer crystals
are typical elemental excitations found in polymeric
systems.
"Conformon" is one of elemental excitations concerning
molecular structure\cite{conformon}.
It is known that some of semi-crystalline polymers form
mesophases that are called a quasi-ordered phase\cite{Yang}.
The phase is defined as a partially disordered crystalline phase,
of which an order parameter is larger than that of amorphous
but smaller than
that of perfect crystal.
One may consider this phase as liquid crystal(LC) with
high viscosity.

Since
the quasi-ordered phase of $\pi$-conjugated polymers is 
closely related to
conjugation length of $\pi$ electrons in a crystal,
it is quite important to investigate both static and dynamic
structure.
In spite of the importance,
information about the dynamics of the phase is still poor,
mainly because there are quite limited experimental methodologies
to approach the problem.
As a theoretical approach, van der Horst and co-workers\cite{J-W} have proposed that in
crystalline conjugated polymers, the presence of static disorder brings about
a spread of the distance between two chains constituting the decrease of
wave-function overlap and the exciton induced by thermally excited intermolecular phonon
modes (dynamic disorder) will reduce the quantum mechanical coherence.
Hence, the conception of a single polymer chain put into a dielectric medium can interpret the features of conjugated polymers.
In this case, the polymers have a character of one-dimensional substances because electrical conduction is dominated by the free electron having only degrees of
freedom which can move in one dimension.

Heat capacity, quasielastic neutron scattering, and nuclear magnetic resonance(NMR)
experiments have
intensively  been performed to explore collective dynamics
of quasi-ordered structure
including
charge and spin density waves\cite{discom},
domain wall dynamics\cite{domain,domain2}
and discommensuration in incommensurate systems\cite{IC}.
So far quasi-ordered phases in polymeric systems are, however, not yet well established.

NMR spectroscopy is one of promissing methods to obtain
information about molecular dynamics since one can perform nondestructive experiments.
Particular attention has been devoted to the phase transiton of $\pi$-conjugated small
molecules which have a class of quasi-ordered phases, the incommensurate(IC) phase (such as biphenyl\cite{biphenyl,Cailleau,Laue}, $\it p$-terphenyl\cite{T_phenyl,p-ter}, bis(4-chlorophenyl)\cite{BCPS} and dichlorobipenylsulphone\cite{Pusiol}, etc.). 
In these phases, disordered crystals show the long range breaking
of the
lattice periodicity.
Therefore the wavelength of the modulation is not an
integral multiple of the
unit cell, which means that translational symmetry
is lost.
In most incommensurate crystals, power-law temperature
and frequency dependences
on proton longitudinal relaxation time were observed.
Furthermore, the relaxation rate rises dramatically
around the transition temperature, T$_{\rm{c}}$.
This is due to the fact that the spin relaxation is
controlled by low-frequency excitations of the collective phase modes (phasons)
corresponding to the critical
slowing down of molecular motion accompanying the
phase transition.

Dynamics of modulation waves in IC solids is also investigated by various
NMR experiments.
In Rb$_2$ZnCl$_4$ (and its analogue)\cite{IC,Blinc_Ailion,Blinc2,papa,papa2,Dolin,Ailion}, bis(4-chlorophenyl) sulfone\cite{BCPS}, and barium sodium niobate\cite{quasi,NQR,Ailion2}, it is known that
the structural modulation wave with respect to atomic displacements from equiblium
positions produces a discommensuration in the higher temperature regime of incommensurate-commensurate transition temperature.
As far as $\pi$-conjugated polymers are concerned, there
are, however, quite limited
publications concerning quasi-ordered phases.
In {\it trans}-polyacetylene one-dimensional diffusion of electron spin along
the chain has been observed\cite{Holc,Holc2,p_ace1,Greene,soliton,SSH2}.
This result could be described by using a well-known soliton (domain-wall) model
which indicates that in the presence of defects or impurities the mobile spins are
trapped and act as the domain wall.

Poly(thiophene-2,5-diyl)[PTh], poly(pyrrole-2,5-diyl)[PPyrr],
and their derivatives (e.g. poly ( 3 - alkylthiophene -2,5- diyl [P3RTh] ),
which are
made up of recurring five-membered rings, have been extensively investigated\cite{handbook,yama1,yama2,yama3,Wudl,McC92,Rieke_jacs,Rieke_synth_metal,Andersson,Rieke_macro,Rieke_jacs95,Holdcroft,Hutchison,P3RTh,yama_jcscc,P4RTz,ChemLett,BullChem,thermochromism2,thermo_p4mtz1,thermo_p4mtz2},
because they are thought to take coplanar structures and
to eventually form highly extended $\pi$-electron conjugated network\cite{Heeger91,Samuelsen,Sam2,Fell} owing to their less sterically hindered structure compared with
those of poly(arylene)s with six-membered rings such as
poly({\it p}-phenylene)[PPP]\cite{ppp}.

PTh and PPyrr are constituted of "electron-excessive"
heterocyclic units
and are susceptible to chemical or electrochemical oxidation, i.e. p-doping.
On the other hand,
it has been recently reported
that $\pi$-electron conjugated polyheterocycles with
five-membered rings
containing "electron-withdrawing" imine nitrogen(s),
poly(4-alkylthiazole-2,5-diyl)[P4RTz],
were synthesized with dehalogenative organometallic poly
condensation\cite{yama_jcscc,P4RTz,HH-P4MeTz_cm}
and have interesting redox and optical properties\cite{LED}.
For solutions of P4MeTz and P3MeTh,
there are no significant differences
in absorption maximum ($\lambda_{\rm max}$) of
UV-visible optical absorption spectra\cite{yama_jacs}; $\lambda_{\rm max}$ = 425nm for the trifluoroacetic acid(TFA) solution of regio-regulated head-to-head(HH)-P4MeTz
and 420 nm for the formic acid solution of {\it rand}-P4MeTz.
For the solvent-cast films of HH-P4MeTz, on the other hand,
the value of $\lambda_{\rm max}$ (498nm) shows
significant red shift compared to
that for the film of {\it rand}-P4MeTz (420nm).
Further,
HH-P4MeTz gives relatively large third-order non-linear
optical susceptibility, $\chi^{(3)}$, of 2.5$\times$10$^{-11}$esu.
The value of $\chi^{(3)}$ is eight times larger than that for {\it rand}-P4MeTz (0.3$\times$10$^{-11}$esu).
These regio-regularity dependence of the optical properties can be observed only for P4MeTz.
There are no similar observations on P3RTh.

\begin{centering}
\begin{figure}
\includegraphics[scale=0.5]{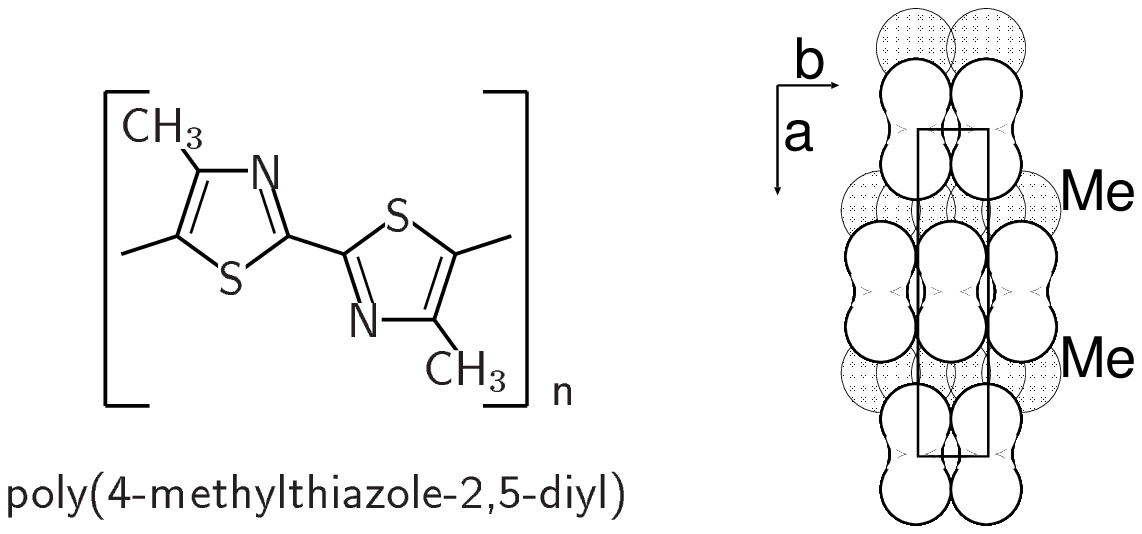}
\caption{Chemical structure and schematic representation of the alternative layered molecular packing
in HH-P4MeTz;
Neighboring molecules take a face-to-face packing (b axis). The X-ray diffraction study shows that thiazole rings recurring along c axis take coplanar structures and form a highly extended $\pi$-conjugated network. 
It takes the alternative layered structure: a layer constituted by highly
densed methyl groups with two dimensional manner and a layer with
the face-to-face $\pi$-stacking.}
\label{alterna}
\end{figure}
\end{centering}

The x-ray diffraction study indicates that
HH-P4MeTz takes a face-to-face $\pi$-stacking structure
while head-to-tail(HT)-P3MeTh takes the staggered $\pi$-stacking\cite{lattice_par,pj}.
One of the interesting structural features of HH-P4MeTz is
an alternative layered structure; one is a layer constituted by
highly densed methyl groups with two dimensional manner
and
the other is built by the face-to-face $\pi$-stacking (Fig.\ref{alterna}).

In this article,
we have synthesized natual abundant and methyl-deuterated (perdeuterated) versions of HH-P4MeTz, and have performed
differential scanning calorimetry(DSC),
variable temperature $^2$D quadrupolar echo NMR,
two-dimensional exchange (2DEX) $^2$D NMR,
variable frequency and temperature proton longitudinal relaxation time (T$_1$),
and variable temperature proton transverse relaxation time(T$_2$) measurements,
in order to explore the dynamics of the quasi-ordered structure of the polymer.
From these experiments, we shall discuss wave length dispersion of modulational waves in the quasi-ordered phase of the polymer and confirmed that there exist the strongly anisotropic diffusion and anomalous dispersion of the structural modulation waves.
\section{Experimental}

\subsection{Materials}
HH-P4MeTz (M$_{\rm w}$=3200; light scattering) was prepared by
the previously reported schemes involving 
dehalogenative organometallic polycondensation\cite{P4RTz}.
A methyl-deuterated version of HH-P4MeTz was synthesized from a scheme initiated from
mono-bromination of deuterated aceton. The detailed scheme for the synthesis will be 
described elsewhere. 
For all the measurements, we used powder samples of HH-P4MeTz, which are recrystallized by hexafluoroisopropanol(HFIP). 

\subsection{Differential scanning calorimetry measurements}
DSC thermograms of natural abundant and perdeuterated versions of HH-P4MeTz were recorded on a Seiko DSC 220 system connected to a SSC5300 workstation.
The samples were first heated from 173K to 423 $\sim$ 473K.
The heating rate were 20 K/min. (HH-P4MeTz) and 5, 20 K/min. (${\it d}_3$-HH-P4MeTz).
After the first heating scans, the samples were rapidly quenched by liquid nitrogen and then they were again heated at the same rate.

\subsection{Solid-state NMR measurements}
\subsubsection{$^1${\rm H} solid state NMR spectroscopy}
We carried out variable frequency proton longitudinal relaxation time (T$_1$)
measurements in order to investigate the spectral density function of fluctuation of the local field in HH-P4MeTz.
The proton longitudinal relaxation times were measured using following three methods;
the saturation-recovery method(T$_{1\rm{H}}$) for three Larmor frequencies of
25(0.59 Tesla),270(6.34 Tesla),
and 400MHz(9.4 Tesla),
the longitudinal relaxation time in the rotating frame of a radio frequency
field(T$_{1\rho}$)\cite{T1rho},
and in the dipolar order built up by the Jeener-Broekaert method(T$_{1\rm{D}}$)\cite{J-B} were measured at a resonance frequency of 400MHz.
For T$_{1\rho}$ measurements on a wide range of radio-frequency(rf) field, we made a home-built probe with a solenoid type microcoil with a radius of 1.2mm and a length of 2.5mm.
The intensity of rf field was in the range from about 43kHz to 1.37MHz.

The proton transverse relaxation time(T$_2$) measurements were performed at a resonance frequency of 400MHz.
The two-pulse Hahn echo (TPHE) and the Carr-Purcell Meiboom-Gill(CPMG) methods can be used to
observe diffusion-like motions under the condition of local magnetic field gradient
in a crystal\cite{Askw_T2}.
The experiments were performed as a function of echo time.
If there is no diffusion, the apparent T$_2$ by CPMG shows a fixed value independent on
the echo time.
On the contrary, when diffusive motion exists the value of T$_2$ is changed with a
variation of echo time.

\subsubsection{$^2${\rm D} NMR spectroscopy}
We used a standard quadrupolar echo pulse sequence for $^2$D NMR measurements.
The rf-field intensity was set at 64kHz.
The recycle delay for all the experiments was 30s.
Signal averagings were performed with 32-256 scans.
The echo time for quadrupolar echo measurements was set at 0.020ms.

For investigation of dynamic processes such as chemical exchange, cross relaxation,
spin diffusion, and atomic hopping motion, the two-dimensional exchange spectroscopy has
a number of advantages over the one-dimensional techniques.
The 2DEX NMR method particularly provides the valuable
data when trying to observe slow dynamics in solids\cite{Ailion,Ailion2,ernst,Dolin2,Spiess}.
A spatial motion of certain atom (or molecule) such as diffusion or reorientation
is detected in the form of frequency change during the mixing time,
t$_{\rm{mix}}$.
Moving nuclei experience a change of resonace frequency between at the beginning and at the
end of t$_{\rm{mix}}$ and then create cross peaks, while static nuclei without changing
the frequency (or moving nuclei with a frequency of much higher than 1/t$_{\rm{mix}}$) produce the diagonal signals in the 2D spectrum.
\subsubsection{$^{13}${\rm C} NMR spectroscopy}
$^{13}$C cross-polarization and magic-angle sample spinning(CPMAS) NMR measurements were
performed on a JEOL GSX-270 FT NMR spectrometer
(6.34 Tesla).
Solid-state two dimensional spin-echo(2DSE) $^{13}$C CPMAS NMR measurements
were performed
on a Varian Unity400 FT NMR spectrometer (9.4 Tesla)
equipped with a double resonance tunable MAS probe.
The rf field intensities during the Hartmann-Hahn
cross-polarization
were set at 50kHz for the $^{1}$H and $^{13}$C channels.
A contact time for the cross-polarization was 2.0ms.
The intensity of CW proton decoupling was 64kHz.
The recycle delay for all the experiments was 5s.
The MAS speed was monitored and controlled by a personal computer
with optical fibers.
Principal components of chemical shift tensors for methyl carbons
were determined with 2DSE spectroscopy,
which is originally developed by Kolbert $\it et \ al$.\cite{Kolbert}
 and applied to determination of CST components with small CSA
by Asakawa $\it et \ al$.\cite{Askw_mrc}.
A pulse sequence of (preparation) $ - t_1/2 - \pi - t_1/2 - $acq($t_2$) was
used for 2DSE measurements.
The detailed setting up for the 2DSE experiments were the following;
the MAS speed was set at $\omega_{\rm r}/2 \pi$ = 1025Hz $\pm$ 5Hz,
128 $t_1$ values were collected, with 512 acquisitions per $t_1$ value,
and the time increment in the $t_1$ dimension was 122$\mu$s
(the increment for each $t_1/2$ duration was set at the rotor cycle of 1/16).
\section{Theoretical}
\subsection{Spectrum Simulation}
A NMR spectral simulation program for 
2DSE
was written by C language and 
were performed on an IBM-AT compatible personal computer
with using a GNU C compiler.
Powder averagings were performed with random orientations 
with respect to the external magnetic field.
The best fit simulation was picked out by monitoring the value of $\epsilon$,
which is defined as 
\begin{eqnarray}
\varepsilon &=& 1 - \frac{(\sum_m \sum_n J_{mn} I_{mn})^2}
{\sum_m \sum_n J_{mn}^2 \sum_m \sum_n I_{mn}^2}, \nonumber 
\end{eqnarray}
where $J_{\rm{mn}}$ and $I_{\rm{mn}}$ are the experimental and simulated 
signal intensity for the (m,n)th spinning sideband.
To describe a chemical shielding tensor, 
we used the span($\Omega$) and skew($\kappa$)\cite{mason} 
as well as the $\delta$ and $\eta$. 

\subsection{Chemical Shielding Calculation}
$^{13}$C chemical shielding tensors of 
HH-P4MeTz 
were calculated by 
the {\it ab initio} self consistent field(SCF) coupled Hartree-Fock method 
with gauge invariant atomic orbitals(SCF-GIAO)\cite{giao,giao2} and
the second-order M\o ller-Plesset GIAO(MP2-GIAO)\cite{giao-mp2}.
Tail-tail(TT)-bi(4MeTz)
were employed as model compounds for all the calculations
and were optimized by the MP2 method with the 6-31G(d) basis set, 
and the SCF-GIAO and MP2-GIAO shielding calculations were carried out with 
6-31G basis set.
All the {\it ab initio} chemical shielding calculations were
performed with Gaussian 98(Rev.A7) program package\cite{g98}
run on a Cray C916/12256 super computer 
at the Computer Center, Tokyo Institute of Technology,
and SGI Origin2000 or Fujitsu VPP2800 
at Institute for Molecular Science in Okazaki, Japan.
\section{Results and Discussion}
\subsection{DSC measurements}
\begin{centering}
\begin{figure}
\includegraphics[scale=0.5]{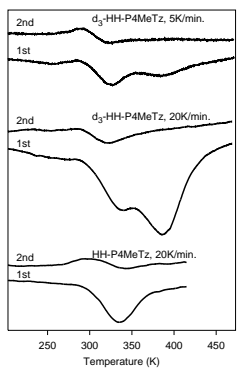}
\caption{DSC charts for powder samples of perdeuterated HH-P4MeTz and natural abundant HH-P4MeTz. The heating rate is shown in the figure.
After the first heating scan, the samples were quenched by liquid nitrogen.}
\label{DSC}
\end{figure}
\end{centering}

The thermal property of natural abundant and perdeuterated HH-P4MeTz were examined
by DSC (Fig.\ref{DSC}).
The thermograms indicate that for both the samples there is an endo-thermal peak with
a specific heat jump near 300K($T_{\rm{c}}$).
This phase transition is due to an order-disorder phase transition, namely, partial melting of
face-to-face $\pi$-stacking.
The similar result was observed in an analog of P4RTz\cite{thermo_p4mtz2}.
Furthermore, another endo-thermal peak appeared at higher temperature ($>$350K) for perdeuterated
HH-P4MeTz.
The phase transition at higher temperature is not due to recrystallization because
an increase of heat of fusion was observed with an increase of heating rate.
It may be due to deuteration induced phase transition,
but one needs further investigation in order to clarify this phenomenon.
In this article, we shall give attention only to the phase transition near 300K.
\subsection{$^2$H Quadrupolar Echo}
\begin{centering}
\begin{figure}
\includegraphics[scale=0.4]{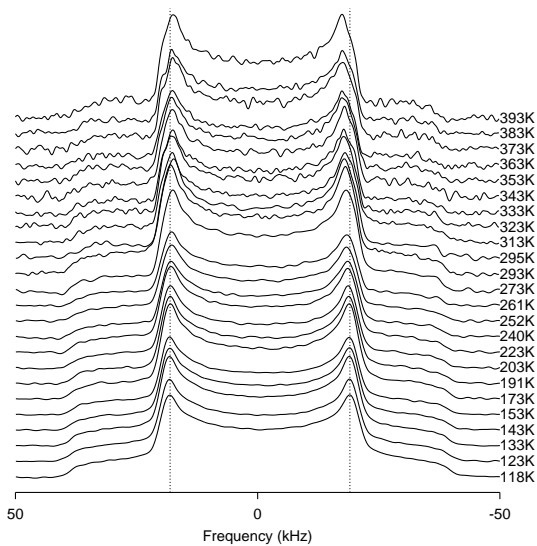}
\caption{41MHz $^2$D duadrupolar echo NMR spectra for the powdered sample of perdeuterated HH-P4MeTz at various temperatures (118 $\sim$ 393K).
The dashed line is only a guide for eyes.
The axially symmetric spectra for the methyl deuterium with C$_{3v}$ rotation were observed below 300K.
Over 300K, the shoulder peaks were detected near the parpendicular edge singularities.}
\label{quad_all}
\end{figure}
\end{centering}

 In order to clarify existence of the quasi-ordered phase,
the $^2$H quadrupolar NMR were performed for perdeuterated HH-P4MeTz at various
temperatures (118-393K).
For the lower temperature phase, in Fig.\ref{quad_all},
the axially symmetric character was observed on
the quadrupolar powder spectra for the methyl deuteriums.
On the other hand, the behavior of edge singularities over the transition temperature is substantially different from those often found in melting of polymers;
no averaging (for cases of commonly observed melting, very sharp peak at the center
of spectrum) was observed in the $^2$H quadrupolar powder pattern and
existence of the axially asymmetric character or another quadrupolar coupling
tensor
was observed.
This feature was observed particularly at the perpendicular edge of the spectra
with a shoulder peak.
So far, several mechanisms have been pointed out to explain the origin of
such a slight change of $^2$H quadrupolar echo spectra.

First,
Hiyama $\it et \ al$. have claimed that
the electrostatic effect in amino acids causes the quadrupolar coupling tensor
of deuterium
to be asymmetric\cite{Hiyama}.
It is difficult to explain our results by the electrostatic effect,
because the effect could be more pronounced in the lower temperature experiments,
whereas no asymmetry character was observed in the low temperature experiments
of HH-P4MeTz.
Second,
Schwartz $\it et \ al$. have pointed out that the
$^2$H-$^2$H magnetic dipolar interaction renders the $^2$H spectrum asymmetric\cite{Schwartz}.
The effect of dipolar interaction can be ruled out
because of the same reason for the electrostatic effect.
Third,
the asymmetric character is also explained by
breaking of C$_{3v}$ symmetry, which have been pointed out by Wann $\it et \ al$.\cite{Wann}.
From Landau's theory of phase transition,
this possibility can be ruled out,
because a symmetry group for the higher temperature phase should belong to a
subgroup
of a symmetry group for the lower temperature phase.
Fourth,
Kintanar $\it et \ al$. have shown that the spectrum can be affected by wagging motion and
resulted in showing the asymmetry\cite{Kintanar}.
From the density measurement and the x-ray diffraction study,
motion of the methyl groups could be substantially hindered if exists.
However, if partial melting of $\pi$-stacking is present, namely,
if a quasi-ordered structure is there,
the nutation will occur on the methyl group in association with backbone twist.
In such a case,
the two signals derived from the crystalline and the quasi-ordered phases
can be observed.
From the above discussion,
it is thought that 
the possibility of the asymmetric quadrupole tensor can be ruled out and
the two methyl sites with slightly different quadrupolar coupling tensor
were observed.

\subsection{Two dimensional Exchange NMR}
\begin{centering}
\begin{figure}
\includegraphics[scale=0.35]{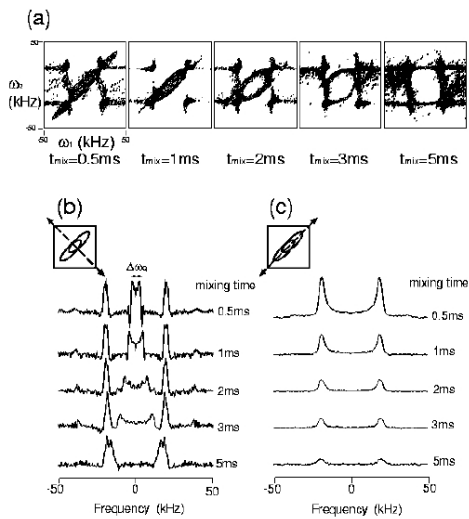}
\caption{The two-dimensional exchange $^2$D NMR spectra for HH-P4MeTz at 288K with various mixing time (a),
The cross-diagonal (b) and the diagonal (c) slice of the 2DEX spectra for HH-P4MeTz.
Tightly fixed or rapidly moving methyl groups will maintain intensities of the diagonal peaks if exist. In our measuring time scale, such the maintainance was not observed, indicating the most methyl groups undergo the slow motion associated with backbone twist.}
\label{2dexch5}
\end{figure}
\end{centering}

\begin{centering}
\begin{figure}
\includegraphics[scale=0.5]{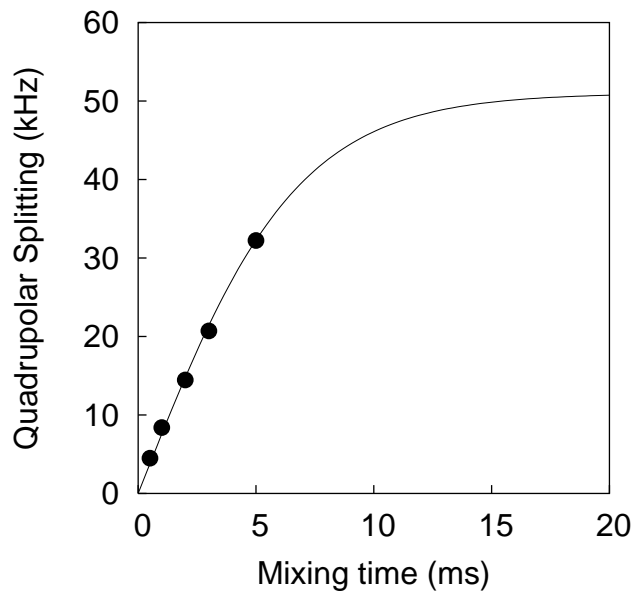}
\caption{The variation of the maximum frequency difference, $\Delta\omega_{\rm Q}$, with mixing time (t$_{\rm {mix}}$) for the non-diagonal peaks. The dotted line is a theoretical curve of Eq.\ref{2dexch}.}
\label{2dtanh}
\end{figure}
\end{centering}

2DEX NMR was performed in order to explore more detailed
molecular dynamics.
Figure \ref{2dexch5}(a) shows the 2DEX NMR spectra for perdeuterated HH-P4MeTz
at 288K as a function of t$_{\rm{mix}}$.
The maximum frequency width between cross peaks ($\Delta \omega_{\rm Q}$) is
increased continuously with the increase of t$_{\rm{mix}}$ (Fig.\ref{2dexch5}(b)).
It was shown that this corresponds to the increase of the
reorientation angle of the methyl group and there exists the motion of
an order of milliseconds in the crystal.
The frequency width $\Delta \omega_{\rm Q}$ is plotted as a function of t$_{\rm{mix}}$ in Fig.\ref{2dtanh}.
Assuming that there is symmetrical two-site exchange,
the correlation time($\tau_{\rm{ex}}$) can be determined from the frequency
width between cross peaks in the 2DEX NMR spectra by
fitting the plot to the following equation\cite{ernst,Spiess}:
\begin{eqnarray}
\Delta \omega_{\rm Q} &=& \rm{A} tanh(\frac{\rm{t}_{\rm mix}}{\tau_{ex}}), \label{2dexch}
\end{eqnarray}
where A is a fitting parameter.
Using this equation the correlation
time is found with $\tau_{\rm{ex}}$ = 6.7ms ($\tau_{\rm{ex}}^{-1}$ = 147Hz)
at 288K.

The similar 2DEX NMR spectrum could be also obtained
by the effect of spin-diffusion between deuteriums instead of molecular motion\cite{Bloem}.
This effect is produced by the flip-flop term between the spins
which consist of magnetic dipole-dipole coupling.
In order to estimate the spin diffusion time $\tau_{\rm{SD}}$
for the zero quantum transition,
the following equation was considered;
\begin{eqnarray}
\frac{1}{\tau_{\rm{SD}}} &=& \frac{1}{\sqrt{2\pi}}
{\mid}{\cal{H}}_{\rm ij} {\mid}^2 L^{\rm{ZQT}} (\omega \rightarrow 0)
\end{eqnarray}
${\mid}{\cal{H}}_{\rm{ij}}{\mid}^{2}$ is the matrix element
of the dipolar flip-flop term and a Lorentzian shape is assumed for the resonance
lines.
Indeed, the line width experiments by CPMG analog for deuterium shows a Lorentzian
shape with a FWHM of $\Delta f$ = 10 Hz.
Each term is given by
\begin{eqnarray}
{\mid}{\cal{H}}_{\rm ij} {\mid}^2 &\sim& \frac{9}{4}\gamma^4 \hbar ^2
\frac{(1-3\rm{cos}\theta_{ij})^2}{r_{\rm{ij}}^6}
\end{eqnarray}
where $r_{\rm{ij}}$ is a typical distance between inter-methyl deuteriums and $\theta_{\rm{ij}}$ is the angle
between the {\bf r}$_{\rm{ij}}$ direction
and magnetic field H$_0$.
The value of $\mid {\cal H}_{\rm{ij}} \mid^2$ is estimated as 560Hz$^2$ by using a typical
distance between the deuteriums of 0.3nm.
\begin{eqnarray}
L^{\rm{ZQT}}(\omega \rightarrow 0) &=& \lim_{\omega \rightarrow 0} \frac{(\rm{T}_2^{\rm{ZQT}})^{-1}}{(\rm{T}_2^{\rm{ZQT}})^{-2}+\omega^2} \nonumber\\
&\sim& \rm{T}_2^{\rm{ZQT}} = \frac{1}{2\pi\Delta \it f}
\end{eqnarray}
Here, $\Delta f$ is assumed to be equal to the homogeneous line width(10 Hz).
$L^{\rm ZQT}(\omega \rightarrow 0)$ is estimated as 15ms.
 Eventually, spin diffusion time, $\tau_{\rm{SD}}$,  is estimated as 0.3s.
This value of $\tau_{\rm{SD}}$ is about 40$\sim$50 times as large as
$\tau_{\rm{ex}}$ (= 6.7ms) observed in the measurements.
Therefore these arguments confirm that the dominant effect which influences
our 2DEX NMR spectra is not spin diffusion but molecular motion.

The diagonal signals reflect the quantity of immobile methyl groups or 
mobile groups with much faster than $\tau_{\rm mix}^{-1}$.
The signal intensity on a diagonal, however, diminishes and disappears with increase in t$_{\rm mix}$ (Fig.\ref{2dexch5}(c)), indicating that no rigid or too rapidly nutating methyl group compared with t$_{\rm mix}$ exists and most of them undergo very slow motions at 288K.
\subsection{Proton Longitudinal Relaxation Time}
\begin{centering}
\begin{figure}
\includegraphics[scale=0.75]{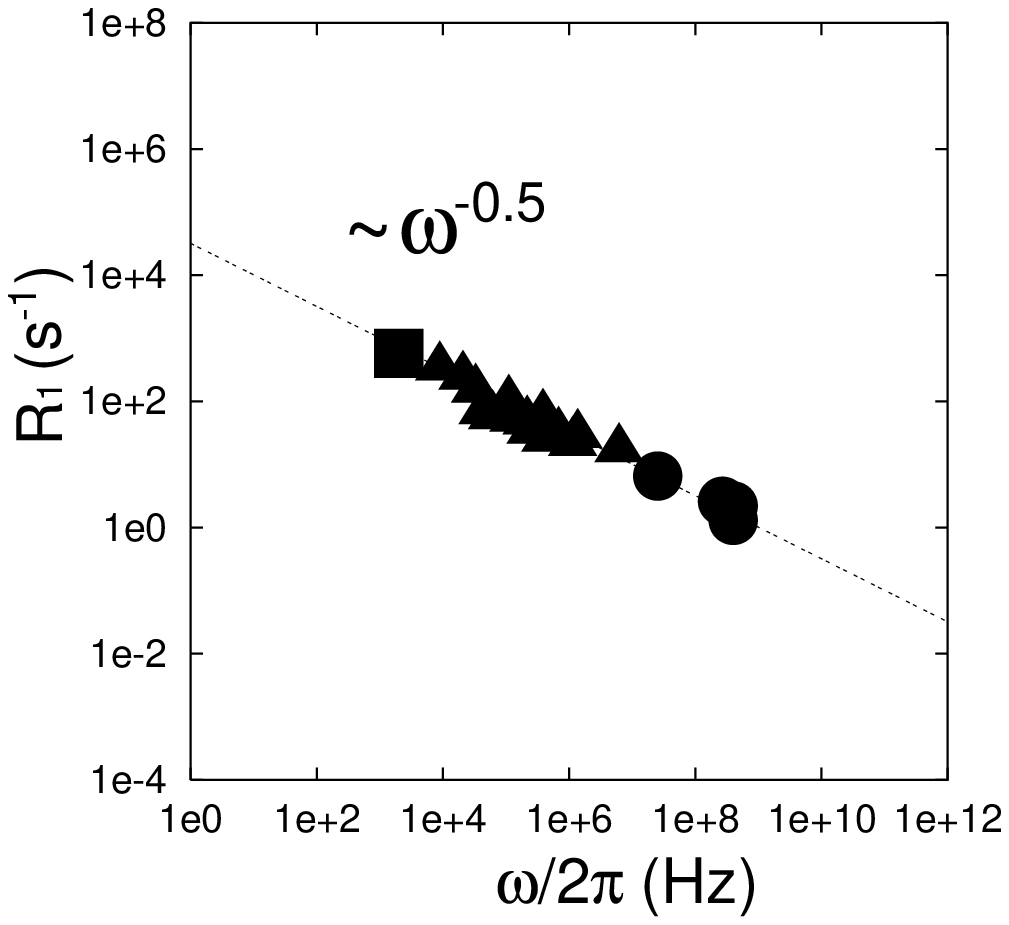}
\caption{The frequency dependence of proton longitudinal relaxation rate for HH-P4MeTz at 288K.
Each symbol shows the results of three different measurement techniques: the saturation-recovery method (closed circle), the spin-locking method (closed triangle) and the Jeener-Broekaert method (closed square) respectively.
The dotted line is the relation of R$_1 \propto \omega^{-1/2}$.
The wide range of frequency of the rotating frame was carried out by using a probe with a micro coil.
The frequency of the dipolar order was determined by the Van Vleck's second moment method.}
\label{tanuki}
\end{figure}
\end{centering}

We performed variable frequency proton spin-lattice relaxation time measurements.
The proton spin-lattice relaxation rate (T$_1^{-1}$ = R$_1$) as a function of the
frequency is shown in Fig.\ref{tanuki}.
The relaxation rate shows the dependence of R$_1$ $\propto$ $\omega^{-1/2}$.
This relationship between R$_1$ and $\omega$ is in disagreement with that predicted
from the classical BPP theory\cite{BPP}.
The BPP theory in which a fluctuation of local field is described as Lorentzian
gives the following spectral density function,
\begin{eqnarray}
J(\omega) &=& \frac{\tau_{\rm c}}{1+(\omega\tau_{\rm c})^2}\ \ 
\sim \ \ \frac{1}{\omega^2\tau_{\rm{c}}} \ \ ( \rm{if} \ \ \omega\tau_{\rm{c}} \gg 1).
\end{eqnarray}
In the region where T$_1$ depends on frequency ($\omega\tau_{\rm{c}}\gg$ 1), the relation R$_1$ $\propto$ $\omega^{-2}$ should be realized.
Therefore an alternative correlation function should
be considered in order to explain the dependence obtained by our experiments.

So far there are several reports that show the $\omega^{-1/2}$ dependence of R$_1$.
The ideas of theories are roughly divided into two categories.
First, with assuming existence of an incommensurate phase,
appropriate dynamic susceptibility of the classical damped harmonic oscillator type
for phason branch leads to the $\omega^{-1/2}$ dependence of spin-lattice relaxation
rate\cite{Blinc,Zeyher}.
At the same time, the theory also derives positive proportionality of
R$_1$ to temperature.
Then, this theory can be ruled out, because a decrease of R$_1$ was observed with an increase of temperature as shown
in Fig.\ref{R1}(a).

The second theory describes the dynamic susceptibility based on a one-dimensional
random walk model (details of this model are given by the reference).
More generally, for the weak collision limit ($\omega\tau_c \ll $ 1), the frequency dependence of R$_1$ is different for one-, two-, and three-dimensional diffusive motions\cite{Conradi,Conradi2},
\begin{eqnarray}
{\rm 1D}: {\rm R}_1 &=& A\tau_c^{1/2}\omega^{-1/2}\nonumber\\
{\rm 2D}: {\rm R}_1 &=& B\tau_c{\rm ln}(\omega)\\
{\rm 3D}: {\rm R}_1 &=& C\tau_c - E\tau_c^{3/2}\omega^{1/2}.\nonumber
\end{eqnarray}
The proportionality constants A, B, C and E depend on the particular model of hopping motion on a given network of atomic sites.
It is clear from Fig.\ref{tanuki} that one-dimensional diffusive motion exists in HH-P4MeTz at our measuring frequency.
The following equation is obtained as a conclusion;
\begin{eqnarray}
{\rm R}_1 &=& {\rm M}_2f(\omega), \label{R1}\\
f(\omega) &=& \frac{1}{\sqrt{2}}\tau_c^{1/2}\omega^{-1/2} \ (\omega < \tau_c^{-1}), \label{fx}
\end{eqnarray}
where $\tau_c$ is the correlation time identical to the inverse of the diffusion rate and M$_2$ is the second moment of the interaction that affects the relaxation.
By fitting the Eq.\ref{R1} to the result of Fig.\ref{tanuki}, the diffusion rate is found to be 3$\pm$2 GHz.
Surprisingly, the rate is an order of 7 larger than the exchange rate obtained
from the 2DEX NMR measurement.

Since longitudinal relaxation time measurements are sensitive to molecular motion
with smaller correlation time than the 2DEX NMR,
it may not be surprising even if the two methods (2DEX NMR {\it vs}. T$_1$) show the huge difference in the correlation time.

In the case of undoped $\it{trans}$\ -polyacetylene, one-dimensional electron hopping affects longitudinal relaxation time and diffusion rate is found to be an order of 10$^{12}$$\sim$10$^{13}$ Hz at room temperature by applying the one-dimensional random walk model.
It is difficults in the present study to estimate the second moment
by taking into account electron spin-spin interaction and hyperfine contribution,
because an amount of free electron per unit cell that is proportional to the second moment is not known.
If mobile electrons govern the T$_1$ measurements, the relaxation rate should become large with increase in temperature\cite{Holc2}.
However, as shown in Fig.\ref{R1}(a), such dependence was not observed for HH-P4MeTz.
Furthermore, we observed large attenuation of
signal intensity at diagonal positions in the 2DEX NMR spectra(see above),
which implies absense of possible high frequency motions sensitive
to longitudinal relaxation time measurements.

Taking into account of absence of high frequency motion in the 2DEX measurements and one-dimensional fluctuation in proton T$_1$ measurements, we can reach
one of plausible candidates to explain the huge difference in the correlation
time, which is wave length dispersion of the modulation waves.
The distribution of wave number produces the localized wave packet by spatial superposition.
In materials with anomalous dispersion, localized waves propagate more rapidly than invdividual waves.
It can be pronounced that the 2DEX NMR is sensitive to the local motion of
backbone twist, namely, the nutation of individual methyl group was observed (which has a character of phase velocity)
and that the T$_1$ measuments are sensitive to one-dimensional
diffusive motion, namely the localized modulation waves (group velocity).

\begin{centering}
\begin{figure}
\includegraphics[scale=0.35]{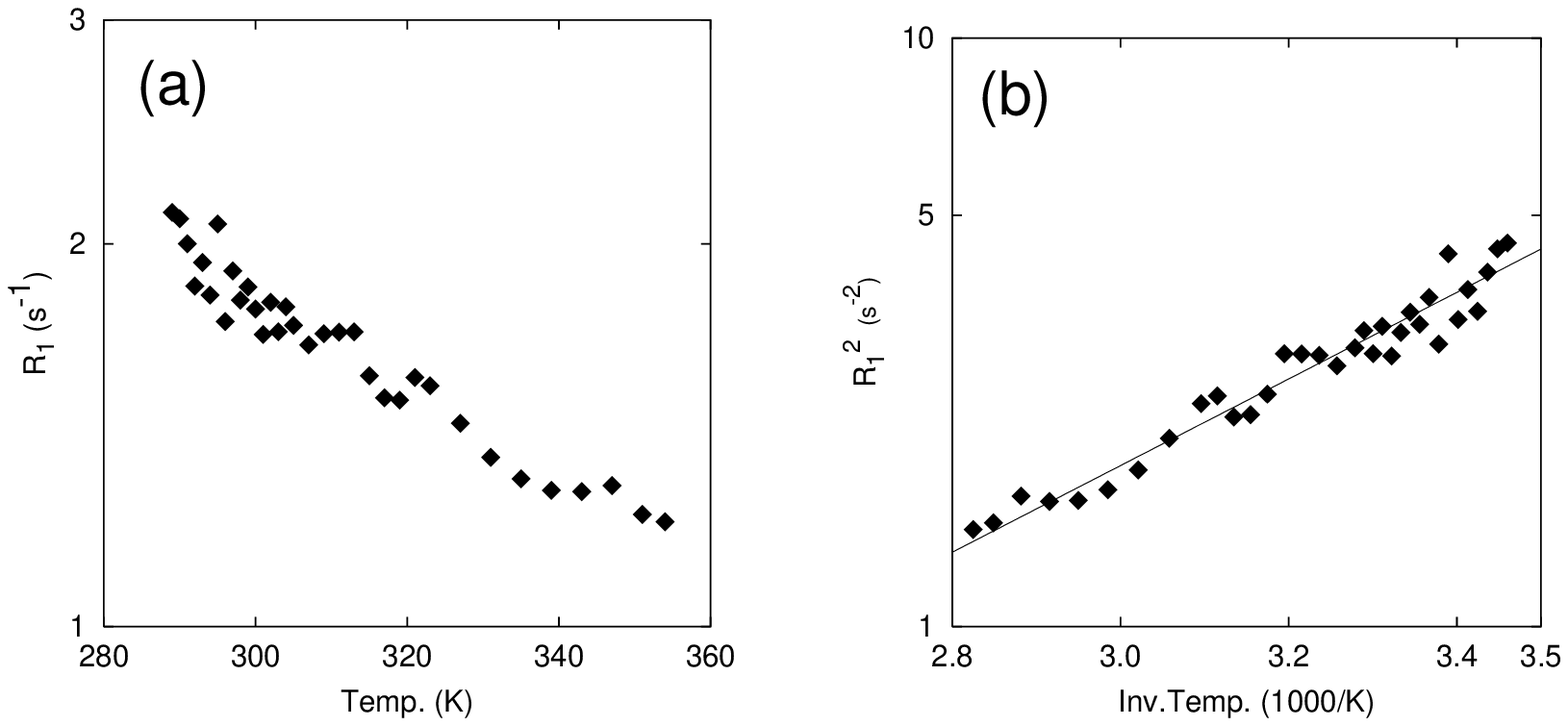}
\caption{The temperature dependence of proton relaxation rate over the temperature range from 290 to 355K.
The value of R$_1$ becomes smaller with elevation of temperature (a).
The plot of R$_1^{2}$ as a function of 1000/T (b).
When one-dimensional diffusion is responsible for the behavior of R$_1$, R$_1^{2}$ is proportional to $\tau_c$ (see Eq.\ref{T1_arrhe}).
The thermally activated process with an Arrhenius type was able to be observed.}
\label{temp_rate}
\end{figure}
\end{centering}

If the relaxation process is governed by thermally activated molecular motions with an Arrhenius type, we can apply the following expressions;
\begin{eqnarray}
\tau_c^{-1} &=& \tau_0^{-1} {\rm exp}(-\frac{E_a}{kT})
\end{eqnarray}
and if the diffusive motion keeps one-dimensionality in our measuring temperature range, from Eqs.\ref{R1} and \ref{fx},
\begin{eqnarray}
(R_1)^2 &\propto& \tau_c = \tau_0 {\rm exp}(+\frac{E_a}{kT}) \label{T1_arrhe}
\end{eqnarray}
is valid.
As Fig.\ref{temp_rate}(b) shows, the correlation time follows the Arrhenius law above the phase transition temperature and the activation energy is found to be 3.4 kcal/mol.
\subsection{Proton Transversal relaxation time}
\begin{centering}
\begin{figure}
\includegraphics[scale=0.4]{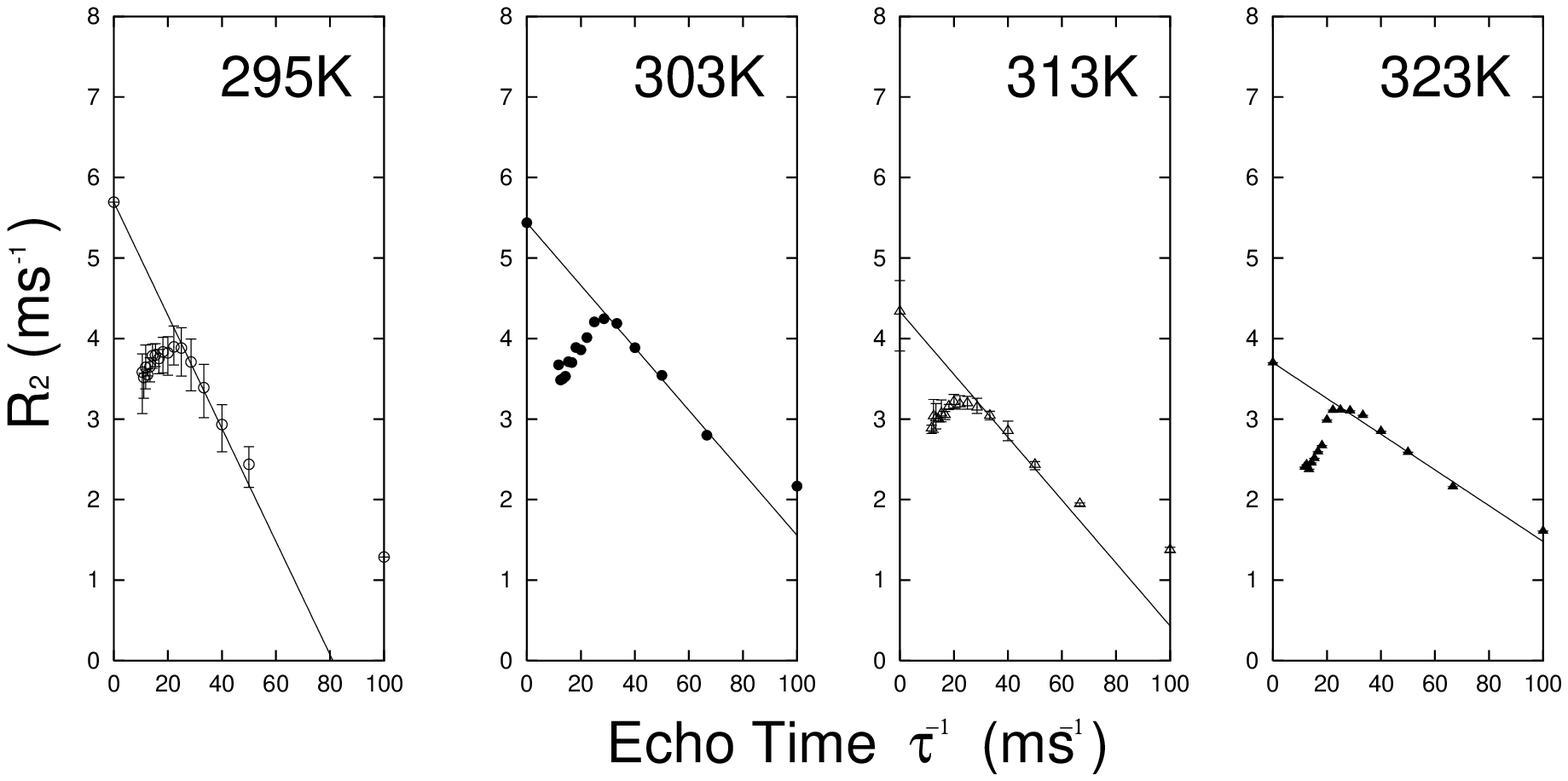}
\caption{The dependence of transverse relaxation rate on CPMG echo time.
The data for two-pulse Hahn echo(TPHE) were added to the plots since the TPHE method is thought as a CPMG experiment with infinite echo time.
The solid lines denotes the extrapolation of the transverse relaxation rate at the short time regime to the relaxation rate of TPHE.}
\label{T2}
\end{figure}
\end{centering}

\begin{centering}
\begin{figure}
\includegraphics[scale=0.35]{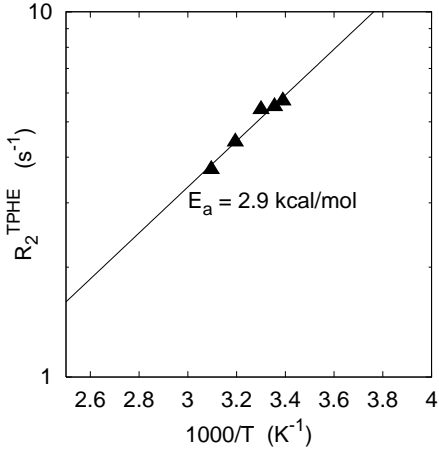}
\caption{The variation of R$_2$ by TPHE as a function of temperature near above T$_{\rm c}$. The thermally activated process was observed and the activation energy was found to be 2.9 kcal/mol.}
\label{T2_arrhe}
\end{figure}
\end{centering}

\begin{centering}
\begin{figure}
\includegraphics[scale=0.65]{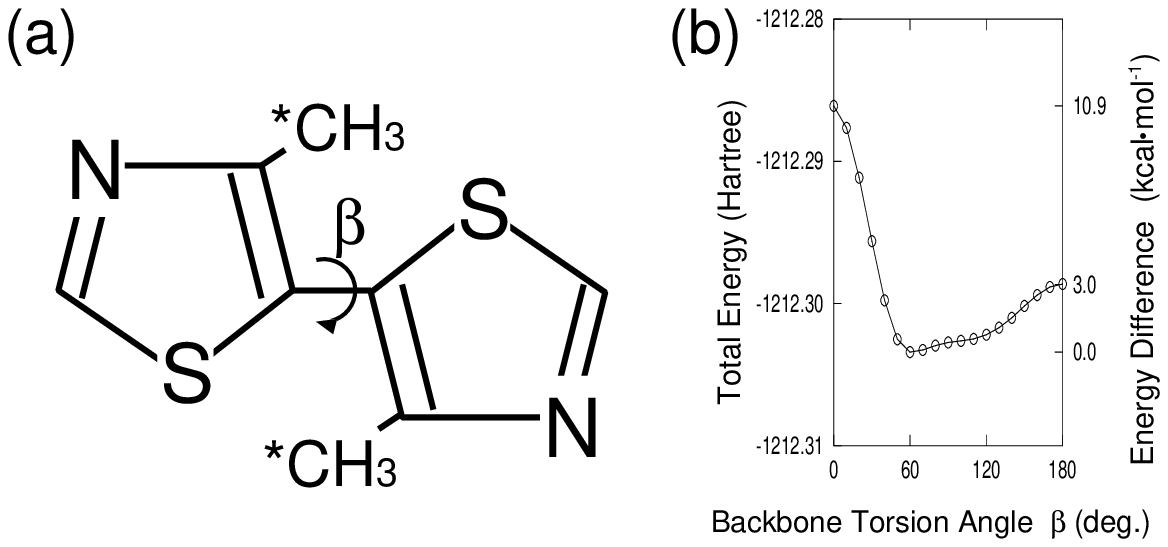}
\caption{A model compounds for quantum mechanical chemical shielding calculation (a). Chemical shieldings
for the carbons marked with an asterisk were calculated.  The two methyl carbons
in TT-P4MeTz is electronically equivalent.
The MP2 energy for HH-P4MeTz as a function of backbone twist of $\beta$ (b).}
\label{model}
\end{figure}
\end{centering}

We report the results of proton transverse relaxation time measurements in order to
verify whether the diffusive motion really exists in
HH-P4MeTz.
One can gradually incorporate the effect of diffusion into transverse relaxation
time measured with CPMG method with an increase of echo time($\tau$).
In the result of CPMG measurements in HH-P4MeTz (Fig.\ref{T2}),
we observed the gradual increase of T$_2^{-1}$(= R$_2$) for smaller $\tau$.
The result of TPHE measurements was added to the plot since this method is thought of as a CPMG method with infinite
echo time (R$_2^{\rm TPHE}$ $\approx$ $\lim_{\tau \rightarrow \infty}$R$_2^{\rm CPMG}$).
The decay function of TPHE for HH-P4MeTz was exponential with first order\cite{Askw_T2}
and
the apparent transverse relaxation time, R$_2^{\rm TPHE}$, is much larger than that of the CPMG measurements.
From the arguments by Robertson\cite{Robertson}, and Le Doussal and Sen\cite{Doussal}, it is known that when these two conditions are realized, the restricted diffusion model can be applied;
\begin{eqnarray}
R_2^{{\rm TPHE}} &\propto&  D^{-1} \label{Robertson}
\end{eqnarray}
For larger $\tau$ ($>$ 40$\mu$s) we also observed the decrease of R$_2$  with an increase of $\tau$.
The behavior of R$_2$ can be understood by Sen's theory which treats a restricted diffusion of magnetization by Bloch-Torrey equation\cite{Sen}.
From the theory,
the short time regime ($\tau < 40\mu$s) is dominated by the diffusion process.
For the larger echo time,
the diffusion of the modulation waves reached at boundaries
(ex., pinning of modulation waves due to the defects or interphase between crystal or amorphous
and the quasi-ordered structure), and
the relaxation behavior corresponds to the localization regime,
where no diffusive character was observed.
Therefore the diffusive motion in HH-P4MeTz was confirmed by
proton transverse relaxation time measurements.
Considering the interchain face-to-face $\pi$-stacked structure,
we can attribute the diffusion not to a class of molecular diffusion like chain diffusion, but to
diffusion of the conformational modulation waves related to backbone twist.
The behavior of the temperature dependence of R$_2^{\rm TPHE}$
obeys Arrhenius law.
Therefore the following relation is led from Eq.\ref{Robertson},
\begin{eqnarray}
R_2^{{\rm TPHE}} &\propto& D^{-1} \ = \ D_0^{-1} {\rm exp}(+\frac{E_a}{kT}),
\end{eqnarray}
and the activation energy was found to be
2.9 kcal/mol (Fig.\ref{T2_arrhe}).
Moreover, {\it ab initio} molecular orbital calculation with the MP2/6-31G(d) method of
TT-4-methylthiazole dimer (geometry optimized by the same basis set) shows
the activation energy of 3.0 kcal/mol for the backbone twist (Fig.\ref{model}).
\subsection{$^{13}$C CPMAS and 2D Spin-echo NMR spectroscopy}
\begin{centering}
\begin{figure}
\includegraphics[scale=0.5]{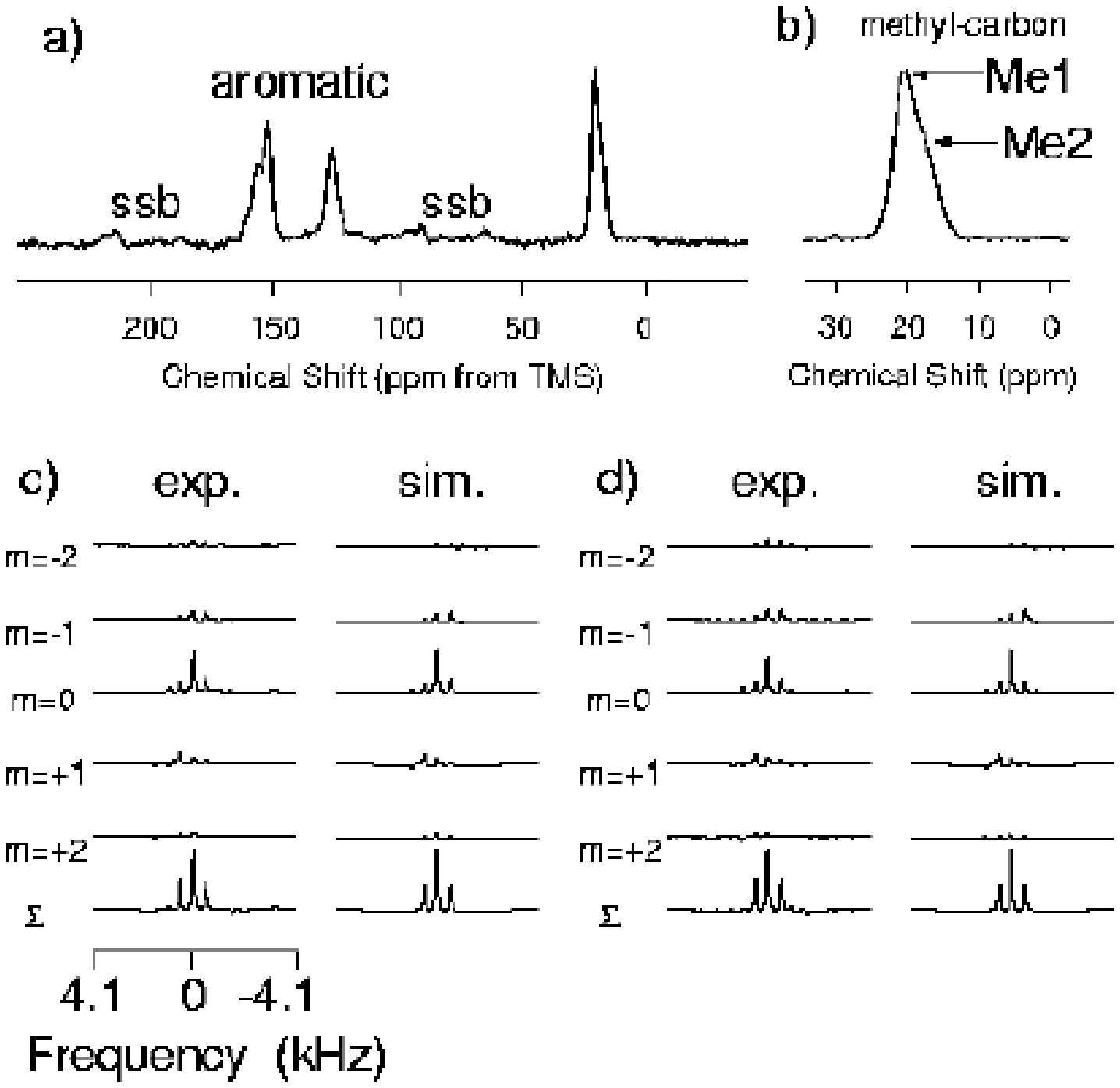}
\caption{67.9MHz $^{13}$C cross-polarization and magic-angle sample spinning(CPMAS) spectra for HH-P4MeTz at 293K(a,b).
(c),(d) 100.0 MHz observed and best fitted simulated
solid-state $^{13}$C two dimensional spin-echo CPMAS
spectra for the methyl carbons in HH-P4MeTz(Me$_1$;c) and HH-P4MeTz(Me$_2$;d).}
\label{CPM}
\end{figure}
\end{centering}

Figure\ref{CPM}(a) and (b) show the $^{13}$C cross-polarization and magic-angle sample
spinning(CPMAS) spectra for HH-P4MeTz at 293K.
For the spectrum,
only signals derived from methyl carbons appear over the shift region
of 15-21 ppm.
It is worth to point out that
the shoulder signal of the methyl carbon
are observed at 17 ppm.
Although the previous X-ray diffraction study\cite{HH-P4MeTz_cm}
suggests that there exists an electronically unique methyl carbon
in the unit cell of HH-P4MeTz,
the NMR spectrum indicates that there exist at least two electronically
distinct methyl carbons.

Chemical shielding tensor(CST) is more informative than its isotropic shielding, about three-dimensional electronic
structure.
It is difficult, unfortunately, to determine principal values of CST
for nuclei with small CSAs, such as methyl carbons.
In order to gain further insights on the chemical shielding difference,
we performed solid-state $^{13}$C two-dimensional spin-echo(2DSE) CPMAS
measurements
and extract the information about principal values of CST
for the methyl carbons.
Fig.\ref{CPM}(c) and (d) shows the solid-state 2DSE spectra
and the best fitted simulated spectra for
the methyl carbons in HH-P4MeTz.
The values of error parameter, $\epsilon$, were determined as
8.3$\times$10$^{-3}$ for the Me$_1$ of HH-P4MeTz,
3.4$\times$10$^{-2}$ for the Me$_2$ of HH-P4MeTz,
respectively.
Table \ref{exp_shielding} summarized the observed principal
components of chemical shielding tensor for the methyl carbons in HH-P4MeTz.
It is realized that the chemical shielding anisotropies for
the two distinctive signals of the methyl carbons
in HH-P4MeTz are quite different from
each other.
Each principal components for the Me$_1$ signal is less shielded
compared to those for the Me$_2$ signal.
This can be mostly pronounced for the midfield component, $\delta_{22}$;
the difference is approximately 6ppm.
This dominates the other principal components and gives the isotropic shielding
difference of 67\%.
We found that the inhomogeneously broadened signals of the methyl carbon
is due to rearrangements of the backbone torsion
in the crystalline lattice, namely, due to coexistence of the quasi-ordered phase with the crystalline phase,
in which the polymer chains stack in a manner
similar to the crystalline polymers but
the $\pi$-$\pi$ interactions are weaker.
With the consideration of the results of 2DEX, the quasi-ordered phase might dynamically coexist with the crystalline phase.

The experimental results of $^{13}$C CST is further understood by {\it ab initio} chemical shielding calculation.
In Table \ref{cry_amor} an averaging of the shielding tensor with respect to $\beta$ was carried out under the consideration of the Boltzmann factor.
The experimental chemical shielding for the methyl carbon in HH-P4MeTz is qualitatively reproduced by variation of backbone twist in the quasi-ordered phase.

\section{Conclusion}
Solid-state dynamics of regio-regulated HH-P4MeTz was investigated by various solid-state NMR spectroscopy.
DSC measurements show that both natural abundant and perdeuterated HH-P4MeTz causes the phase transition near 300K.
In the $^2$D quadrupolar NMR spectra at various temperatures
also suggest that the phase transition from the crystalline to the quasi-ordered phase occurs.
The two-dimensional exchange NMR experiments revealed that the motion of methyl groups with an order of milliseconds exists in HH-P4MeTz and
the exchange rate of such the motion was estimated as $\tau_{\rm ex}^{-1}$ = 147 Hz.
Furthermore, the relationship of R$_1$ $\sim$ $\omega^{-1/2}$ was observed from proton longitudinal relaxation time measurements, which might be due to the one-dimensional diffusion-like motion of conformational modulation waves related to the nutation of methyl groups along the chain.
Existence of diffusion of modulation waves was also confirmed from proton transverse relaxation time experiments.
We obtained the diffusion rate of 3$\pm$2 GHz by calculating the dynamic susceptibility with assumption of the 1D random walk model.
The discrepancy between the results of the T$_1$ measurements and 2DEX NMR may be due to the anomalous dispersion of the modulation wave.
$^{13}$C CPMAS NMR measurements,
an analysis of principal components of CST for the methyl carbons determined 
by the 2DSE measurements, 
and chemical shielding calculation show 
that there exists a quasi-ordered phase in HH-P4MeTz and 
that the backbone twist is highly correlated with the methyl nutation.

From the $^2$D 2DEX NMR and the $^1$H T$_1$ measurements, it is also found that HH-P4MeTz has an anomalous dispersion of the modulation wave.
This research is the first example that have measured phonon dispersion by NMR.
Such dynamics may be detectable due to existence of distinctive one-dimensional fluctuation in $\pi$-conjugated systems.

\newpage
\begin{centering}
\begin{table}
\caption{Principal components of
observed nuclear shielding tensors
of the methyl carbons in HH-P4MeTz (units in parts per million from tetramethylsilane.)}
\begin{tabular}{lccccccccc}
\hline
\hline
material &  $\delta_{\rm iso}$ & $\delta_{11}$ & $\delta_{22}$ & $\delta_{33}$ & $\delta$ & $\eta$ & $\Omega$$^a$ & $\kappa$$^b$ & $\epsilon$ \\
\hline
HH-P4MeTz(Me$_1$) & 20.1 & 29 & 23 & 8 & -12 & 0.5  & 21 & 0.41  & 8.3$\times$10$^{-3}$\\
HH-P4MeTz(Me$_2$) & 17   & 28 & 17 & 6 & -11 & 1.0  & 22 & 0.00  & 3.4$\times$10$^{-2}$\\
\hline
\hline
\end{tabular}\\
a)\ span: $\Omega = \delta_{11} - \delta_{33}$ \\
b)\ skew: $\kappa =3(\delta_{22} - \delta_{\rm iso})/(\delta_{11} - \delta_{33})$
\label{exp_shielding}

\newpage
\caption{Principal components of
calculated absolute nuclear shielding tensors
of methyl carbons in HH-P4MeTz (units in parts per million.)}
\begin{tabular}{lccccccccc}
\hline
\hline
model &  Method & $\sigma_{\rm iso}$ & $\sigma_{11}$ & $\sigma_{22}$ & $\sigma_{
33}$ & $\delta$ & $\eta$ & $\Omega$ & $\kappa$ \\
\hline
s-{\it trans} & SCF & 186.7 & 178.2 & 180.4 & 201.4 & -14.7 & 0.150 & 23.2 & 0.8
14\\
averaged      & SCF & 189.8 & 180.3 & 184.5 & 204.4 & -14.6 & 0.288 & 24.1 & 0.6
60 \\
s-{\it trans} & MP2 & 190.6 & 181.2 & 184.4 & 206.3 & -15.7 & 0.204 & 25.1 & 0.7
41 \\
averaged      & MP2 & 194.2 & 184.1 & 188.7 & 209.8 & -15.6 & 0.295 & 25.7 & 0.6
42 \\
\hline
\hline
\end{tabular}
\label{cry_amor}
\end{table}
\end{centering}

\clearpage

\begin{thebibliography}{999}
\bibitem{conformon}
E.R.Andrew,
Phys.Letts. {\bf 34A}, 30 (1971).
\bibitem{Yang}
C.Yang, F.P.Orfino, and S.Holdcroft,
Macromolecules {\bf 29}, 6510 (1996).
\bibitem{J-W}  
J.-W. van der Horst, P.A.Bobbert, and M.A.J.Michels,
Phys.Rev.B {\bf 64}, 035206 (2002).
\bibitem{discom}
W.L.Mcmillan,
Phys.Rev.B {\bf 14}, 1496 (1976).
\bibitem{domain}
K.S.Novoselov, A.K.Geim, S.V.Dubonos, E.W.Hill, and I.V.Grigorieva,
Nature {\bf 426}, 812 (2003).
\bibitem{domain2}
T.Janssen, O.Radulescu,
Z.Phys.B {\bf 104}, 657 (1997).
\bibitem{IC}
R.Blinc, David C.Ailion, J.Dolinsek, and S.Zumer,
Phys.Rev.Lett. {\bf 54}, 79 (1985).
\bibitem{biphenyl}
S.-B.Liu and M.S.Conradi,
Phys.Rev.Lett. {\bf 54}, 1287 (1985).
\bibitem{Cailleau}
J.Etrillard, B.Toudic, H.Cailleau, and G.Coddens,
Phys.Rev.B {\bf 51}, 8753 (1995).
\bibitem{Laue}
L.von Laue, F.Ermark, A.G$\ddot{\rm o}$lzh$\ddot{\rm a}$user, U.Haeberlent, and U.H$\ddot{\rm a}$cker,
J. Phys.:Condens.Matter {\bf 8}, 3977 (1996).
\bibitem{T_phenyl}
K.Kohda, N.Nakamura, and H.Chihara,
J. Phys. Soc. Japan {\bf 51}, 3936 (1982).
\bibitem{p-ter}
J. O.Williams,
Chem.Phys.Lett. {\bf 42}, 171 (1976).
\bibitem{BCPS}
R.Blinc, U.Mikac, T.Apih, J.Dolinsek, J.Seliger, J.Slak, S.Zumer, L.Guibe, and D. C.Ailion,
Phys.Rev.Lett. {\bf 88}, 015701 (2002).
\bibitem{Pusiol}  
R.E. de Souza, M.Engelsberg, D.J.Pusiol,
Phys.Rev.Lett. {\bf 66}, 1505 (1991).
\bibitem{Blinc_Ailion}
R.Blinc, D.C.Ailion, P.Prelovsek, and V.Rutar,
Phys.Rev.Lett. {\bf 50} 67 (1983).
\bibitem{Blinc2}
R.Blinc, F.Milia, B.Topi$\check{\rm c}$, and S.$\check{\rm Z}$umer,
Phys.Rev.B {\bf 29}, 4173 (1984).
\bibitem{papa}
F.Milia, G.Papavassiliou, and A.Anagnostopoulos,
Phys.Rev.B {\bf 43}, 11464 (1991).
\bibitem{papa2}
G.Papavassiliou, A.Leventis, and F.Milia,
Phys.Rev.Lett. {\bf 74}, 2387 (1995).
\bibitem{Dolin}
J.Dolinsek and G.Papavassiliou,
Phys.Rev.B {\bf 55}, 8755 (1997).
\bibitem{Ailion}
L. Muntean and D. C.Ailion,
Phys.Rev.B {\bf 62}, 11351 (2000).
\bibitem{quasi}
J. A.Norcross, D. C.Ailion, R.Blinc, J.Dolinsek, T.Apih, and J.Slak,
Phys.Rev.B. {\bf 50}, 3625 (1994).
\bibitem{NQR}
D. C.Ailion and J. A.Norcross,
Phys.Rev.Lett. {\bf 74}, 2383 (1995).
\bibitem{Ailion2}
L. Muntean and D.C.Ailion,
Phys.Rev.B {\bf 63}, 012406 (2000).
\bibitem{Holc}
K.Holczer, J.P.Boucher, F.Devreux, and M.Nechtschein,
Phys.Rev.B {\bf 23}, 1051 (1981).
\bibitem{Holc2}
M.Nechtschein, F.Devreux, F.Genoud, M.Guglielmi, and K.Holczer,
Phys.Rev.B {\bf 27}, 61 (1983).
\bibitem{p_ace1}
B.R.Weinberger, E.Ehrenfreund, A.Pron, A.J.Heeger, and A.G.MacDiarmid,
J.Chem.Phys. {\bf 72}, 4749 (1980).
\bibitem{Greene}
M.Nechtschein, F.Devreux, R.L.Greene, T.C.Clarke, and G.B.Street,
Phys.Rev.Lett. {\bf 44}, 356 (1980).
\bibitem{soliton}
W.P.Su, J.R.Schrieffer, and A.J.Heeger,
Phys.Rev.Lett. {\bf 42}, 1698 (1979).
\bibitem{SSH2}
W.P.Su, J.R.Schrieffer, and A.J.Heeger,
Phys.Rev.B {\bf 22}, 2099 (1980).
\bibitem{handbook}
R.D.McCullough, {\it Handbook of Oligo- and Polythiophenes}
(Wiley-VCH, Weinheim, 1999).
\bibitem{yama1}
T.Yamamoto, K.Sanechika, and A.Yamamoto, 
J. Polym. Sci. Polym.Lett.Ed. {\bf 18}, 9 (1980).
\bibitem{yama2}
T.Yamamoto and K.Sanechika,
Chem. Ind. (London) 301 (1982).
\bibitem{yama3}
T.Yamamoto, T. Maruyama, H. Suganuma, M. Arai, D. Komarudin, and S. Sakai,
Chem.Lett. 139 (1997).
\bibitem{Wudl}
R.M.Souto Maior, K.Hinkelmann, H.Eckert, and F.Wudl,
Macromolecules {\bf 23}, 1268 (1990).
\bibitem{McC92}
R.D.McCullough and R.D.Lowe,
J. Chem. Soc. Chem. Commun. 70 (1992).
\bibitem{Rieke_jacs}
T.-A.Chen, X.Wu, R.D.Rieke,
J. Am. Chem. Soc. {\bf 117}, 233 (1995).
\bibitem{Rieke_synth_metal}
T.-A. Chen, R.D.Rieke,
Synth. Met. {\bf 60}, 175 (1993).
\bibitem{Andersson}
M.R.Andersson, D.Selse, M.Berggren, H.J$\ddot{\rm a}$rvinen
T.Hjertberg, O.Ingan$\ddot{\rm a}$s, O.Wennerstr$\ddot{\rm o}$m, and
J.-E.$\ddot{\rm O}$sterholm,
Macromolecules {\bf 27}, 6503 (1994).
\bibitem{Rieke_macro}
X.Wu, T.-A. Chen, and R.D.Rieke,
Macromolecules {\bf 28}, 2101 (1995).
\bibitem{Rieke_jacs95}
T.-A. Chen, X.Wu, and R.D.Rieke,
J.Am.Chem.Soc. {\bf 117}, 233 (1995).
\bibitem{Holdcroft}
M.I.Arroyo-Villan, G.A.Diaz-Quijada,
M.S.A.Abdou, and S.Holdcroft,
Macromolecules {\bf 28}, 975 (1995).
\bibitem{Hutchison}
S.C.Rasmussen, J.C.Pickens, and J.E.Hutchison,
Chem. Mater. {\bf 10}, 1990 (1998).
\bibitem{P3RTh}
Y.Miyazaki and T.Yamamoto,
Synth. Met. {\bf 64}, 69 (1994).
\bibitem{yama_jcscc}
T.Yamamoto, H.Suganuma, T.Maruyama, and K.Kubota,
J. Chem. Soc. Chem. Commun. 1613 (1995).
\bibitem{P4RTz}
T.Maruyama, H.Suganuma, and T.Yamamoto,
Synth. Met. {\bf 74}, 183 (1995).
\bibitem{ChemLett}
T.Yamamoto,
Chem. Lett. 703 (1996).
\bibitem{BullChem}
T.Yamamoto,
Bull. Chem. Soc. Jpn. {\bf 72}, 621 (1999).
\bibitem{thermochromism2}
W.R.Salaneck, O.Ingan$\ddot{\rm a}$s, B.Th$\acute{\rm e}$mans, J.O.Nilsson, B.Sj$\ddot{\rm o}$gren, J.-E.$\ddot{\rm O}$sterholm,nJ.-L.Br$\grave{\rm e}$das, and S.Svensson,
J. Chem. Phys. {\bf 89}, 4813 (1988).
\bibitem{thermo_p4mtz1}
J.I.Nanos, J.W.Kampf, M.D.Curtis, L.Gonzalez, and D.C.Martin,
Chem. Mater. {\bf 7}, N2232 (1995).
\bibitem{thermo_p4mtz2}
L.Gonzalez-Ronda, D.C.Martin, J.I.Nanos, J.K.Politis, and M.D.Curtis,
Macromolecules {\bf 32}, 4558 (1999).
\bibitem{Heeger91}
M.J.Winokur, P.Wamsley, J.Moulton, P.Smith, A.J.Heeger,
Macromolecules {\bf 24}, 3812 (1991).
\bibitem{Samuelsen}
J.M\aa rdalen, E.J.Samuelsen, O.R.Gautun, and P.H.Carlse
n,
Synth. Met. {\bf 48}, 363 (1992).
\bibitem{Sam2}
R.D.McCullough, S.Tristram-Nagle, S.P.Williams,
R.D.Lowe, and M.Jayaraman,
J. Am. Chem. Soc. {\bf 115}, 4910 (1993).
\bibitem{Fell}
H.J.Fell, E.J.Samuelsen, E.Bakken, and P.H.J.Carlsen,
Synth.Met. {\bf 72}, 193 (1995).
\bibitem{ppp}
T.Yamamoto, A.Morita, Y.Miyazaki,
T.Maruyama, H.Wakayama, Z.h. Zhou,
Y.Nakamura, T.Kanbara, S.Sasaki, K.Kubota,
Macromolecules {\bf 25}, 1214 (1992).
\bibitem{HH-P4MeTz_cm}
T.Yamamoto, H.Suganuma, T.Maruyama, T.Inoue, Y.Muramatsu,
M.Arai, D.Komarudin, N.Ooba, S.Tomaru, S.Sasaki, and K.Kubota,
Chem. Mater. {\bf 9}, 1217 (1997).
\bibitem{LED}
T.Yamamoto, H.Suganuma, Y.Saitoh, T.Maruyama, and T.Inoue,
Jpn.J.Appl.Phys. {\bf 35}, L1142 (1996).
\bibitem{yama_jacs}
T.Yamamoto, D.Komarudin, M.Arai, B.-L. Lee, H.Suganuma,
N.Asakawa, Y.Inoue,
K.Kubota, S.Sasaki, T.Fukuda, H.Matsuda.
J. Am. Chem. Soc. {\bf 120}, 2047 (1998).
\bibitem{lattice_par}
HH-P4MeTz has a C-centered orthorhombic unit cell with
a=14.1 \AA, b=3.64 \AA, and c=7.61 \AA(see Ref.\cite{HH-P4MeTz_cm}).
\bibitem{pj}
T.Yamamoto, B.-L.Lee, H.Suganuma, S.Sasaki,
Polymer, J. {\bf 30}, 853 (1998).
\bibitem{T1rho}
D.C.Ailion, {\it in Advances in Magnetic Resonance} edited by J.S.Waugh (Academic press, New York, 1971) {\bf 5}, 177.
\bibitem{J-B}
J.Jeener and P.Broekaert,
Phys.Rev. {\bf 157}, 232 (1967).
\bibitem{Askw_T2}
N. Asakawa, T. Kajikawa, K. Sato, M. Sakurai, Y. Inoue, and T. Yamamoto,
J.Mol.Struct. {\bf 602-603}, 455 (2002).
\bibitem{ernst}
R.R.Ernst, G. Bodenhausen, and A. Wokaun,
{\it Principles of Nuclear Magnetic Resonance in One and Two Dimensions}
(Clarendon Press, Oxford, 1987).
\bibitem{Dolin2}
J.Dolinsek, B.Ambrosini, P.Vonlanthen,
J.L.Gavilano, M.A.Chernikov, and H.R.Ott,
Phys.Rev.Lett. {\bf 81}, 3671 (1998).
\bibitem{Spiess}
K.Schmidt-Rohr and H.W.Spiess,
{\it Multidimensional Solid-state NMR and Polymers}
(Academic press, London, 1994)
\bibitem{Kolbert}
A.C.Kolbert, D.P.Raleigh, M.H.Levitt, and R.G.Griffin,
J. Chem. Phys. {\bf 90}, 679 (1989).
\bibitem{Askw_mrc}
N.Asakawa, M.Takenoiri, D.Sato, M.Sakurai, and Y.Inoue,
Magn. Reson. Chem. {\bf 37}, 303 (1999).
\bibitem{mason}
J.Mason, 
Solid State Nucl. Magn. Reson. {\bf 2}, 285 (1993).
\bibitem{giao}
D.Ditchfield,
Mole. Phys. {\bf 27}, 789 (1974).
\bibitem{giao2}
K.Wolinsky, J.F.Hinton, P.Pulay,
J. Am. Chem. Soc. {\bf 112}, 8251 (1990).
\bibitem{giao-mp2}
Gauss, J.\
Chem. Phys. Lett. {\bf 191}, 614 (1992).
\bibitem{g98}
Gaussian 98, Revision A.7,
M.J.Frisch, G.W.Trucks, H.B.Schlegel, G.E.Scuseria,
M.A.Robb, J.R.Cheeseman, V.G.Zakrzewski, J.A.Montgomery.Jr.,
R.E.Stratmann, J.C.Burant, S.Dapprich, J.M.Millam,
A.D.Daniels, K.N.Kudin, M.C.Strain, O.Farkas, J.Tomasi,
V.Barone, M.Cossi, R.Cammi, B.Mennucci, C.Pomelli,
C.Adamo,
S.Clifford, J.Ochterski, G.A.Petersson, P.Y.Ayala,
Q.Cui,
K.Morokuma, D.K.Malick, A.D.Rabuck, K.Raghavachari,
J.B.Foresman, J.Cioslowski, J.V.Ortiz, A.G.Baboul,
B.B.Stefanov, G.Liu, A.Liashenko, P.Piskorz, I.Komaromi,
R.Gomperts, R.L.Martin, D.J.Fox, T.Keith, M.A.Al-Laham,
C.Y.Peng, A.Nanayakkara, C.Gonzalez, M.Challacombe,
P.M.W.Gill, B.Johnson, W.Chen, M.W.Wong, J.L.Andres,
C.Gonzalez, M.Head-Gordon, E.S.Replogle, and J.A.Pople,
(Gaussian, Inc., Pittsburgh PA, 1998).
\bibitem{Hiyama}
Y.Hiyama, S.Roy, K.Guo, L.G.Butler, and D.A.Torchia,
J.Am.Chem.Soc. {\bf 109}, 2525 (1987).
\bibitem{Schwartz}
L.J.Schwartz, E.Meirovitch, J.A.Ripmeester, and J.H.Freed,
J.Phys.Chem. {\bf 87}, 4453 (1983).
\bibitem{Wann}
W.H.Wann and G.S.Harbison,
J.Chem.Phys. {\bf 101}, 231 (1994).
\bibitem{Kintanar}
A.Kintanar, T.M.Alam, W.C.Haung, C.Schindele, D.E.Wemmer, and G.Drobny,
J.Am.Chem.Soc. {\bf 110}, 6367 (1988).
\bibitem{Bloem}
N.Bloembergen, S.Shapiro, P.S.Pershan, and O.Artman,
Phys.Rev. {\bf 114}, 445 (1959).
\bibitem{BPP}
N.Bloembergen, E.M.Purcell, and R.V.Pound,
Phys.Rev. {\bf 73}, 679 (1948).
\bibitem{Conradi}
A.F.McDowell, C.F.Mendelsohn, M.S.Conradi, R.C.Bowman, and A.J.Maeland,
Phys.Rev.B {\bf 51}, 6336 (1995).
\bibitem{Conradi2}
F.Kimmerle, G.Majer, U.Kaess, A.J.Maeland, M.S.Conradi, and A.F McDowell,
J.Alloys and Compounds {\bf 264}, 63 (1998).
\bibitem{Blinc}
S.$\check{\rm{Z}}$umer and R.Blinc,
J.Phys.C {\bf 14}, 465-484 (1981).
\bibitem{Zeyher}
R.Zeyher and W.Finger,
Phys.Rev.Lett. {\bf 49}, 1833 (1982).
\bibitem{Robertson}
B.Robertson,
Phys.Rev. {\bf 151}, 273 (1966).
\bibitem{Doussal}
P.Le Doussal and P.N.Sen,
Phys.Rev.B {\bf 46}, 3465 (1992).
\bibitem{Sen}
S.Axelrod and P.N.Sen,
J.Chem.Phys. {\bf 114}, 6878 (2001).
\end{thebibliography}
\end{document}